\documentclass[%]
 reprint,
superscriptaddress,
 amsmath,amssymb,
 floatfix,
 aps
]{revtex4-2}

\usepackage{graphicx}
\usepackage{dcolumn}
\usepackage{bm}
\usepackage{caption}
\usepackage{subcaption}
\usepackage{svg}
\usepackage[linkcolor=black,colorlinks=true,citecolor=black,filecolor=black]{hyperref}
\usepackage{comment}
\usepackage{amsmath}
\usepackage{yhmath}
\usepackage{booktabs}
\usepackage{tabularx}
\usepackage{amssymb}

\hypersetup{
    colorlinks=true,
    linkcolor=blue,
    filecolor=magenta,      
    }

\begin{document}

\preprint{APS/123-QED}

\title{Paramagnetic half-moon shaped diffuse scattering arising from 3D magnetic frustration}

\author{Nelly Natsch}
\email{nelly.natsch@gmail.com}
\affiliation{Materials Theory, ETH Zürich, Wolfgang-Pauli-Strasse 27, 8093 Zürich, Switzerland}

\author{Tara N. Tošić}
\affiliation{Materials Theory, ETH Zürich, Wolfgang-Pauli-Strasse 27, 8093 Zürich, Switzerland}

\author{Jian-Rui Soh}
\affiliation{A*STAR Quantum Innovation Centre (Q.InC), Institute of Materials Research and Engineering (IMRE), Agency for Science Technology and Research (A*STAR), 2 Fusionopolis Way, Singapore 138634}

\author{Nicola A. Spaldin}
\email{nspaldin@ethz.ch}
\affiliation{Materials Theory, ETH Zürich, Wolfgang-Pauli-Strasse 27, 8093 Zürich, Switzerland}

\date{\today}

\begin{abstract}
We use spin dynamics simulations to determine the origin of the unusual correlated diffuse scattering, characterised by half-moon shapes bridging the magnetic Bragg peaks, observed in the polarised elastic neutron scattering from manganese tungstate, MnWO\textsubscript{4}. We first fit a Heisenberg Hamiltonian with twelve nearest-neighbour exchange interactions and single-ion anisotropy to the experimental ground-state magnon dispersion. We then show via spin dynamics simulations that our model Hamiltonian both reproduces the experimentally observed half-moon features and captures their persistence into the paramagnetic regime. Moreover, we identify the three-dimensional, competing antiferromagnetic interactions driving this behavior. Our work complements earlier studies of half-moon-shaped signatures in pyrochlore and triangular structures, by providing insight into their origin in a zigzag chain geometry with three-dimensional competing exchange interactions.
\end{abstract}

\maketitle

\section{Introduction}

\begin{figure*}
    \centering
    \includegraphics[width=.95\linewidth]{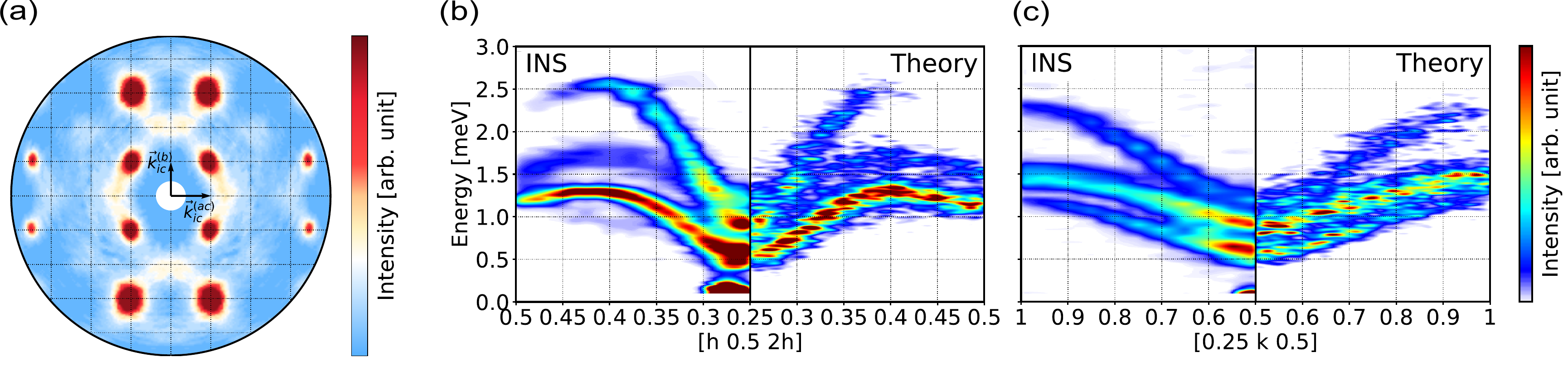}
    \caption{(a) The elastic scattering map at 14 K in the plane spanned by $\vec{k}^{\,(b)}_{ic}$ = [0, -0.5, 0] and $\vec{k}^{\,(ac)}_{ic}$ = [-0.21, 0, 0.45] (fractional coordinates), from Fig. 5.13 in Ref. \cite{holbein_neutron_2016} (reproduced with permission). The signal is purely magnetic, obtained from polarization analysis averaging spin-flip and non spin-flip channels \cite{holbein_neutron_2016}. (b) and (c) AF1 magnon dispersions. The left panels in each plot are INS measurements from Ref. \cite{xiao_spin-wave_2016}. The right panels in each plot show our computed dynamical structure factors $S(\overrightarrow{q}, w)$. (b) and (c) span the [1 0 2] and [0 1 0] directions, respectively.}
    \label{fig:INS}
\end{figure*}

Neutron scattering measurements on a wide range of geometrically frustrated materials show curved, diffuse signals in reciprocal $\vec{q}^{\,}$ space. One prevalent signature is ``split rings of scattering" \cite{yan_half_2018}, also referred to as half moons, found in the metallic spin ice Pr\textsubscript{2}Ir\textsubscript{2}O\textsubscript{7} \cite{udagawa_out--equilibrium_2016} as well as in Kagome classical spin liquids \cite{mizoguchi_clustering_2017}. These dispersive signatures have been variously ascribed to weakly interacting spin loops \cite{mizoguchi_clustering_2017, udagawa_out--equilibrium_2016}, locally ordered domains \cite{paddison_magnetic_2018}, magnons propagating in a cooperative paramagnetic state \cite{bai_magnetic_2019} or a superposition of zero-energy modes \cite{tomiyasu2013emergence}. Such ring-like features in the diffuse signal can be captured by frustrated nearest-neighbor antiferromagnetic Heisenberg models \cite{conlon_absent_2010}. In addition, it has been shown that including weak further-distance neighbors in the Hamiltonian affects the high-temperature paramagnetic signature, e.g. by suppressing pinch points between Bragg peaks \cite{conlon_absent_2010}.\\
\indent While half moons in the scattering from systems with pyrochlore and triangular lattices have been relatively well studied \cite{yan_half_2018, udagawa_out--equilibrium_2016, mizoguchi_clustering_2017, conlon_absent_2010, mizoguchi_2018}, similar signals, observed in the multiferroic MnWO\textsubscript{4}, are less well understood. The magnetic Mn\textsuperscript{2+} ions (S = 5/2) form zigzag chains along the $c$ axis \cite{weitzel_kristallstrukturverfeinerung_1976, holbein_neutron_2016} and frustration arises from competing long-range exchange interactions \cite{xiao_spin-wave_2016}. In MnWO\textsubscript{4}, the half-moon-shaped correlated diffuse scattering is located symmetrically mirrored around $q_a=k_{ic}^{(a)}=-0.21$ and $q_a=3*k_{ic}^{(a)}=-0.63$ (fractional coordinates), bridging the magnetic Bragg peaks \cite{holbein_neutron_2016} (Fig.~\ref{fig:INS}\,a)), and persisting in the paramagnetic regime to up to three times the Néel temperature, $T_\mathrm{N}$ \cite{holbein_neutron_2016}. Here, we shed light on the magnetic interactions in MnWO$_4$ responsible for the appearance of this directional diffuse scattering in the paramagnetic regime.\\
\indent MnWO\textsubscript{4} is paramagnetic at room temperature and undergoes three magnetic phase transitions on cooling in zero magnetic field. The first transition occurs at $T_\mathrm{N}$ = 13.5 K into the AF3 phase \cite{lautenschlager_magnetic_1993}, a collinear, antiferromagnetic alignment of moments along the easy axis $\hat{e}_{ac}$, which lies in the $ac$ plane and forms a 37° angle with the $a$ axis \cite{holbein_neutron_2016}. The order is a sinusoidal modulation with an incommensurate propagation vector $\vec{k}^{\,}_{ic} = [-0.214,$ $ 0.5, 0.457]$ \cite{lautenschlager_magnetic_1993}.
Below 12.3 K, in the AF2 phase, an additional magnetic order develops along the $b$ direction, giving rise to an elliptical spin spiral with the same propagation vector $\vec{k}^{\,}_{ic}$ as in the AF3 phase. The third phase transition occurs at 8.0 K, below which the system orders in the AF1 ground state, in which the moments are again aligned along $\hat{e}_{ac}$, with a commensurate $\vec{k_c}^{\,} = [-0.25, 0.5, 0.5]$ magnetic propagation vector \cite{lautenschlager_magnetic_1993}.

\begin{figure}
    \centering
    \includegraphics[width=0.9\linewidth]{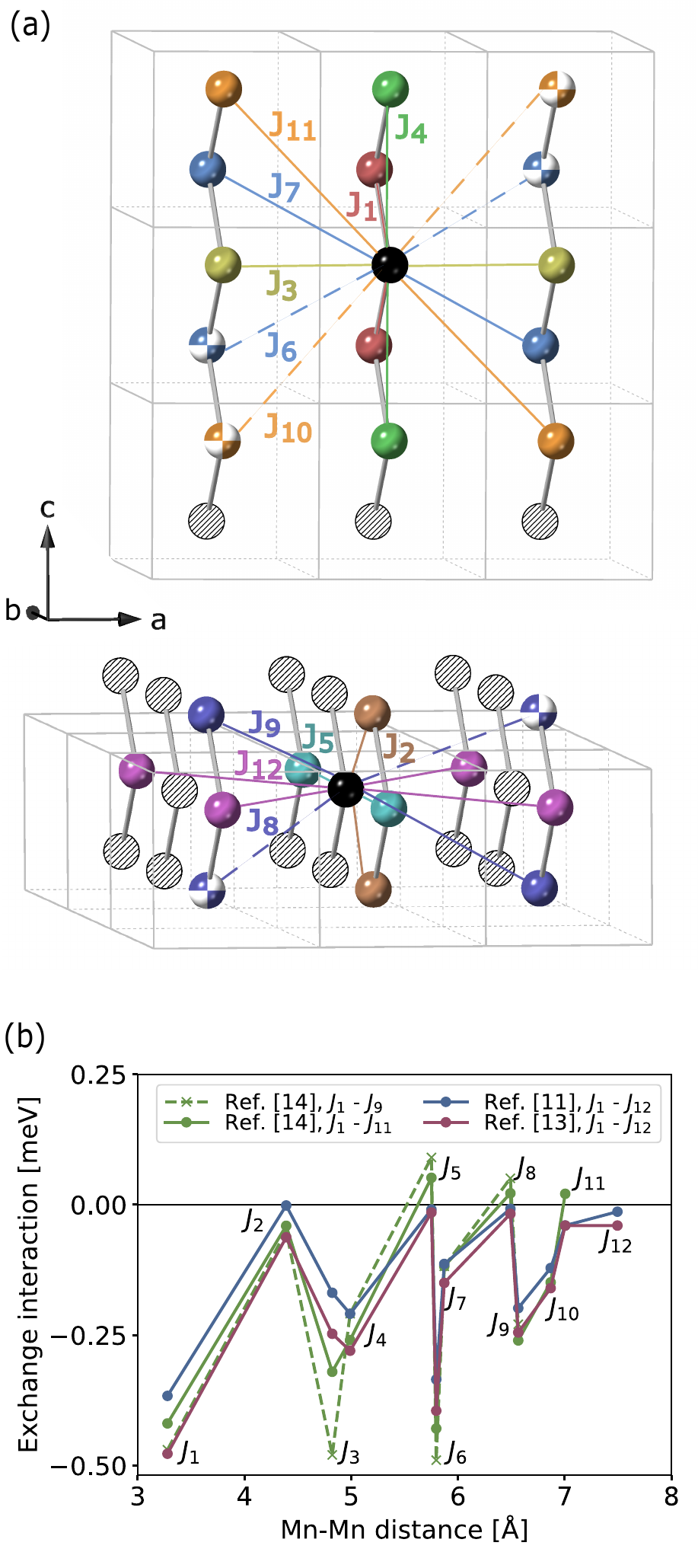}
    \caption{(a) The 12 nearest exchange interactions between the Mn ions. Top panel:
    Couplings within one $ac$ layer of zigzag chains.
    Bottom panel:
    Couplings across different $ac$ layers of zigzag chains.
    We show the interactions of a representative black atom, coloring the neighboring atoms according to the respective exchange interaction. Half-colored atoms indicate that they have a Mn-Mn distance very similar to the respective fully colored atom. The exchange interactions are labeled numerically by order of increasing Mn-Mn distance, following Ref. \cite{xiao_spin-wave_2016}. The grey boxes indicate the structural unit cell. (b) Exchange interactions obtained from Refs. \cite{xiao_spin-wave_2016, wang_comprehensive_2018, ye_long-range_2011} from fitting to INS results.}
    \label{fig:Js}
\end{figure}

The magnetic dynamics of the AF1 phase have been well characterized. We plot the measured inelastic neutron scattering (INS) spectrum of the AF1 phase from Ref. \cite{xiao_spin-wave_2016} in the left panels of Fig.~\ref{fig:INS}\,b) and c), showing the four magnon branches, as well as localized, low-energy signals around $\vec{k_c}^{\,}$ arising from electromagnons. The measured AF1 magnon dispersions from several experimental studies have been described by an extension of the Heisenberg Hamiltonian with a uniaxial anisotropy term \cite{ye_long-range_2011, xiao_spin-wave_2016, wang_comprehensive_2018}:

\begin{equation} \label{eq: H_UppASD}
H =-\sum_{i \neq j} J_{i j} \ \hat{e}_{i} \cdot \hat{e}_{j} + D \sum_i (\hat{e}_{i}\cdot \hat{e}_{ac})^2  \quad .
\end{equation}
Here, $J_{ij}$ are the exchange couplings, as defined in Fig. \ref{fig:Js}\,a), $D$ is the uniaxial single-ion anisotropy, $\hat{e}_{i,j}$ are the normalized magnetic moments, and $\hat{e}_{ac}$ is the direction of the easy axis.

\indent We plot the fitted Hamiltonian parameters from Refs.~\cite{xiao_spin-wave_2016, wang_comprehensive_2018, ye_long-range_2011} in Fig.~\ref{fig:Js}\,b). The initial study in Ref. \cite{ye_long-range_2011} includes 11 exchange interactions, which we label $J_{1}$-$J_{11}$, dropping $ij$ for conciseness. Subsequent studies \cite{wang_comprehensive_2018, xiao_spin-wave_2016} find that the the quality of the fit improves upon adding a twelfth interaction, $J_{12}$, with similar values obtained in Refs. \cite{xiao_spin-wave_2016} and \cite{wang_comprehensive_2018}. Fig. \ref{fig:Js}\,b) shows that several long-range interactions are comparable in strength to the AFM nearest-neighbor interaction $J_{1}$, suggesting a considerable degree of magnetic frustration \cite{xiao_spin-wave_2016}. The origin of the frustration lies in the distorted [MnO\textsubscript{6}] octahedra, which have 95° Mn-O-Mn bond angles \cite{ehrenberg_magnon_1999, holbein_neutron_2016, weitzel_kristallstrukturverfeinerung_1976}.\\
\indent While the magnitude and signs of the exchange interactions are well established, the origin of the half moons in MnWO\textsubscript{4} is not clear. In this work, we aim to capture the short-range correlations giving rise to the half-moon-shaped correlated diffuse scattering by simulating the elastic neutron scattering measured in Ref. \cite{holbein_neutron_2016}. Since previous reports indicate that the half moons in MnWO\textsubscript{4} are not captured by a mean-field model \cite{holbein_neutron_2016}, we employ spin dynamics simulations, which successfully have been shown to capture the magnetic domain-wall structures in the AF2 phase of Co-doped MnWO\textsubscript{4} \cite{leo_polarization_2015}.\\
\indent Our paper is structured as follows. First, we set up an atomistic spin dynamics model with a Heisenberg Hamiltonian that we fit to the experimental AF1 magnon dispersions of MnWO\textsubscript{4}. Next, we calculate structure factors, aiming to reproduce the paramagnetic scattering maps measured in Ref. \cite{holbein_neutron_2016}. Finally, following a similar approach to Ref. \cite{tosic_2024}, we re-calculate the structure factor using simplified Hamiltonians, switching off certain exchange interactions. In this way, we can pinpoint the specific exchange interactions that are responsible for the half moons, and provide insight into the interplay between magnetic frustration and correlated diffuse scattering in MnWO\textsubscript{4}.\\

\section{Methods} \label{sec: methods}
\indent We run atomistic spin dynamics simulations as implemented in the UppASD code (version 6.0.1) \cite{noauthor_uppasd_nodate}, with a damping coefficient of $\alpha$ = 0.05 and 1 femtosecond time steps. The lattice vectors are set to (expressed in Cartesian coordinates and units of \r A):  $\vec{a}^{\,} = [4.824, 0, 0]$, $\vec{b}^{\,} = [0, 5.755, 0]$ and $\vec{c}^{\,} = [-0.095, 0, 5.001]$ \cite{fiz-karlsruhe_entry_2024, urcelay-olabarria_incommensurate_2013}. The two symmetry related manganese atoms occupy the 2$f$ Wyckoff positions, which correspond to $[0.5, 0.685, 0.25]$ and $[0.5, 0.315, 0.75]$ in fractional coordinates \cite{fiz-karlsruhe_entry_2024, urcelay-olabarria_incommensurate_2013}. We set the magnitude of the magnetic moment to 5 $\mu_\mathrm{B}$, since the electrons of the manganese ions are in the high-spin d\textsuperscript{5} configuration \cite{xiao_spin-wave_2016}.\\
\indent For the ground-state magnon dispersions, we calculate the dynamical structure factor $S(\overrightarrow{q},\omega)$ with an energy resolution of $\Delta E = 0.01$ meV, relying on 5000 sampling steps with 83 steps between each sampling. A $48\times48\times48$ supercell is initialized in the AF1 configuration at 1.5 K. We linearly interpolate and use a Gaussian smoothing with a standard deviation of 3.\\
\indent In order to explore the phase transitions, we start from a randomly initialized system and perform Monte Carlo steps from 75\,K to 18\,K, with 20'000 steps performed at 15 log-spaced temperatures. Spin-dynamics calculations, with 30'000 initialization steps and 85'000 measurement steps, are run at 1\,K intervals between 18\,K and 11\,K, and subsequently at 0.25\,K intervals down to 1\,K. \\
\indent To model the correlated diffuse scattering in the paramagnetic phase, we cool the system with spin dynamics steps from 60\,K in 2\,K steps down to 40\,K, followed by 1\,K steps down to 14\,K. At each temperature, we perform 30'000 initialization steps and 40'000 measurement steps. Since Ref.~\cite{holbein_neutron_2016} measures the magnetic elastic scattering, we compute the static structure factor $S(\overrightarrow{q})$, with 40'000 initialization and measurement steps each. Since our aim is to capture the signatures of frustration in the presence of thermal fluctuations within the paramagnetic regime, we compare the experimental and our calculated results at the same absolute temperatures, instead of relative to their respective $T_\mathrm{N}$s. We fit the intensity of the Bragg peaks in our model to the experiment by tuning the post-processing parameters, using a common setting across temperatures. In this way, we set the Gaussian filter to the standard deviation of 4.\\
\vspace{-10pt}
\section{Model development and validation} \label{sec: model development}
\indent We set up our Hamiltonian adopting the relative magnitudes of the exchange interactions and the single-ion anisotropy given in Ref. \cite{xiao_spin-wave_2016}, but finding we have to scale the parameters by a factor (Table \ref{tab:H parameters}). We assume this is due to different ways of incorporating the size of the spin into the exchange interactions and the single-ion anisotropy in different theoretical calculations.\\
\indent We compare our calculated magnon dispersions, obtained by computing the dynamical structure factor $S(\overrightarrow{q},\omega)$, to the INS-measured magnon dispersions in Figs.~\ref{fig:INS}\,b) and c), which show reasonable agreement, confirming the validity of our Hamiltonian. Note that the localized signals at around $[0.25, 0.5, 0.5]$ in the INS-measured spectra arise from electromagnons, which are not captured with Heisenberg Hamiltonians \cite{xiao_spin-wave_2016}.\\
\indent As extracted from the temperature dependence of the heat capacity, we locate a phase transition at around $T_\mathrm{model}$ = 11 K (see Fig.~6 in Supplementary Information). Furthermore, our $T_\mathrm{model}$ = 11 K is reasonably close to the experimental 13.5 K - paramagnetic to AF3 - and 12.3 K - AF3 to AF2 - transition temperatures \cite{lautenschlager_magnetic_1993}. By varying the supercell size and the cooldown speed, our simulation converges to different low temperature configurations, such as the AF1 phase, the AF2 phase and various AF1+spiral phase coexistences (Fig.~\ref{fig:scscizes}  in Supplementary Information). Nevertheless, the AF1 state has the lowest energy, indicating that our model correctly identifies the AF1 phase as the ground state.\\
\indent We point out that, for the purpose of this work, we are not interested in capturing the sequence of the magnetically ordered states. In this context, we note that our model does not reproduce the modulated spin density wave in the AF3 phase, since it does not allow for variations in the magnitude of the magnetic moments.\\
\indent To further validate our model, we test whether our model at 11 K is consistent with the experimental propagation vector $\vec{k}^{\,}_{ic}$. Signals at $\vec{k}^{\,}_{ic}$ are likely to persist in the paramagnetic regime and its associated correlated diffuse scattering. We compute $S(\overrightarrow{q})$ across the entire Brillouin zone, and determine the propagation vector from the $\vec{q}^{\,}$ point with maximum intensity (Fig.~\ref{fig:Sq_fullcellSlices} in Supplementary Information). We find for the 36$\times$24$\times$36 and 48$\times$24$\times$24 supercells $\vec{k}^{\,}_{calc} = [-0.222, 0.5, 0.444]$ and $\vec{k}^{\,}_{calc} = [-0.229, 0.5, 0.458],$ respectively, which are both in reasonable agreement with the experimental $\vec{k}^{\,}_{ic} = [-0.214, 0.5, 0.457]$.\\

\begin{figure*}[t]
\begin{subfigure}[b]{.96\textwidth}
    \centering
    \includegraphics[width=1\linewidth]{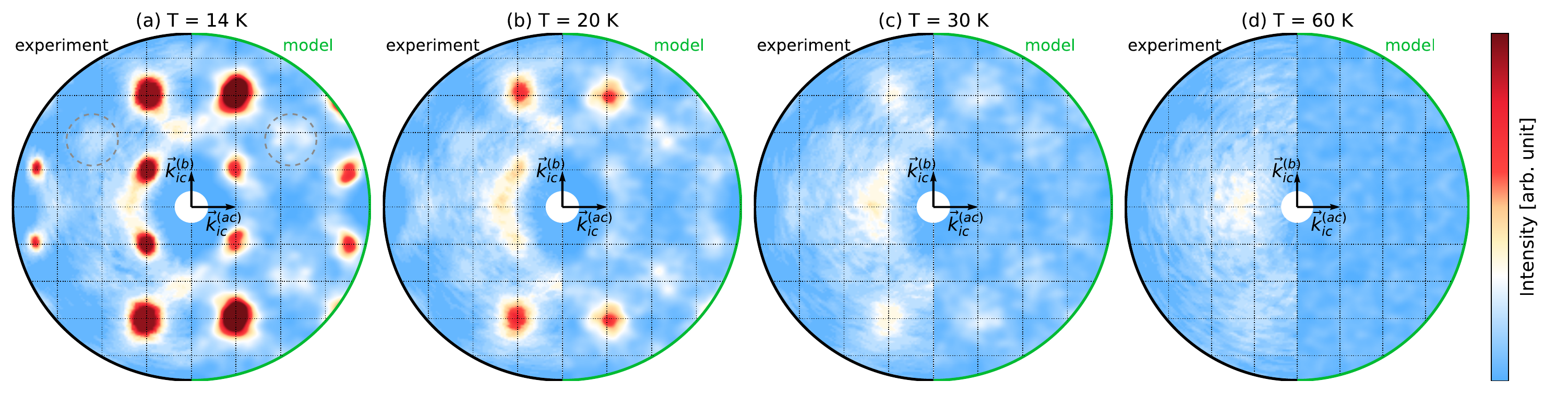}
\end{subfigure}\hspace{5mm}
\caption{Comparison between the experimental scattering maps from Fig. 5.13 in Ref. \cite{holbein_neutron_2016} (reproduced with permission) and our calculated $S(\overrightarrow{q})$, using the Hamiltonian of Eq. \ref{eq: H_UppASD} and the parameters from Table~\ref{tab:H parameters} for the $36\times24\times36$ supercell. 
The reciprocal plane is defined by $\vec{k}^{\,(b)}_{ic}$ = [0, -0.5, 0] and $\vec{k}^{\,(ac)}_{ic}$ = [-0.21, 0, 0.45]. Note that in the case of our model $\vec{k}^{\,(ac)}_{ic}$ = [-0.22, 0, 0.44] (see Fig.~\ref{fig:Sq_fullcellSlices}  in Supplementary Information).}
\label{fig:Sq}
\end{figure*}

\begin{table*}[t]
    \centering
    \caption[The Hamiltonian parameters for UppASD]{The Hamiltonian parameters used in our spin dynamics model, defined by the Hamiltonian of eq. \ref{eq: H_UppASD} and in units of meV. The values are from Ref. \cite{xiao_spin-wave_2016}, but multiplied by 1.6 and -3.2 for the $J$s and $D$ respectively.}
    \begin{tabular}{*{13}{p{12.3mm} }}
    \toprule
     $J_1$ & $J_2$ & $J_3$ & $J_4$ & $J_5$ & $J_6$ & $J_7$ & $J_8$ & $J_9$ & $J_{10}$ & $J_{11}$ & $J_{12}$ & $D$ \\
     \midrule
     -0.592 & -0.0032 & -0.272 & -0.336 & -0.0176 & -0.544 & -0.176 & -0.016 & -0.32 & -0.192 & -0.0672 & -0.0256 & -0.192 \\ 
    \bottomrule
    \end{tabular}

    \label{tab:H parameters}
\end{table*}

\section{Correlated diffuse scattering at \texorpdfstring{T \textgreater T\textsubscript{N}}{TgreaterTN}}
\indent In the following, we show that our spin dynamics model reproduces the paramagnetic correlated diffuse scattering measured in Ref.~\cite{holbein_neutron_2016}. Figs.~\ref{fig:Sq}\,a)-d) compare the computed and experimental scattering maps, showing that the half moons are captured by our model. Furthermore, the computed half moons persist up until 30\,K, whereas only background scattering remains at 60 K, consistent with the experiment of Ref.~\cite{holbein_neutron_2016}. \\
\indent In this context, we note that, in contrast to the simulated $S(\overrightarrow{q})$, the measurements have a significantly higher intensity along the half moon located at $q_a=-0.21$ than along that at $q_a=-0.63$. This is likely due to the anisotropic background caused by the plate-like sample holder \cite{holbein_neutron_2016}, as well as the magnetic form factor \cite{janas_classical_2021, shirane_neutron_2002}. We do not include the magnetic form factor in our model, since we aim to capture collective spin-spin excitations, which are reflected in the structure factor \cite{janas_classical_2021, shirane_neutron_2002}.
It is reasonable to assume that these experimental properties - not captured by our model - are also responsible for the higher measured intensity in the half moon at $q_a=-0.21$ compared to our calculations, in particular at 14\,K. We also note that the intensity distribution and the width of the signals in the simulation depends on computational parameters, such as the number of steps.\\
\indent A set of weak signals (circled in dashed grey in Fig.~\ref{fig:Sq}\,a) is present in both our simulation and the experiment of Ref.~\cite{holbein_neutron_2016}.
Ref.~\cite{holbein_neutron_2016} assigns these signals to $2*\vec{k}^{\,}_{ic}$ and attributes them to exchange striction, with the magnetic modulation inducing a lattice modulation of twice the magnetic propagation vector \cite{taniguchi_magnetic-field_2008}. In fact, the AF2 phase has a $2*\vec{k}^{\,}_{ic}$ lattice modulation and an accompanying $2*\vec{k}^{\,}_{ic}$ magnetic modulation \cite{finger_analyse_2013, finger_polarized-neutron-scattering_2010, biesenkamp_structural_2020}, remnants of which might persist into the paramagnetic regime ~\cite{holbein_neutron_2016}. However, we point out that if the signal were at $2*\vec{k}^{\,}_{ic}$, then it should be centered at the grid line intersection in Fig.~\ref{fig:Sq}\,a. We propose that the signal, found in both our simulation and the measurement of Ref.~\cite{holbein_neutron_2016}, is more likely to sit at the commensurate $2*\vec{k}^{\,}_{c}$. Because of this, and the fact that our model does not include exchange striction, we raise the question of the origin of the weak signal at $2*\vec{k}^{\,}_{c}$ again, and suggest it could arise from a competing or coexisting order.\\
\indent We point out that our model captures the characteristic paramagnetic scattering features, including the half moons, regardless of the supercell size and the simulated low temperature state obtained upon cooling from the paramagnetic state (Fig.~\ref{fig:scscizes}  in Supplementary Information). Therefore, we suggest that the short-range correlated features in the paramagnetic regime are not a consequence of the particular pattern of the long-range magnetic order.\\
\indent We comment that, given S($\vec{q}=0$)=0 in the paramagnetic phase (see Fig.~\ref{fig:zeroslice} in the Supplementary Information) the correlated clusters have no associated magnetic moment. This suggests that the correlated diffuse scattering results from short-range ordered clusters of spins that have zero magnetic moment. Further insight would be gained by studying the response of the diffuse scattering to an external magnetic field, which would couple to a net magnetic moment.\\

\section{Simpler Hamiltonian models}

\begin{figure}
    \centering
    \includegraphics[width=.95\linewidth]{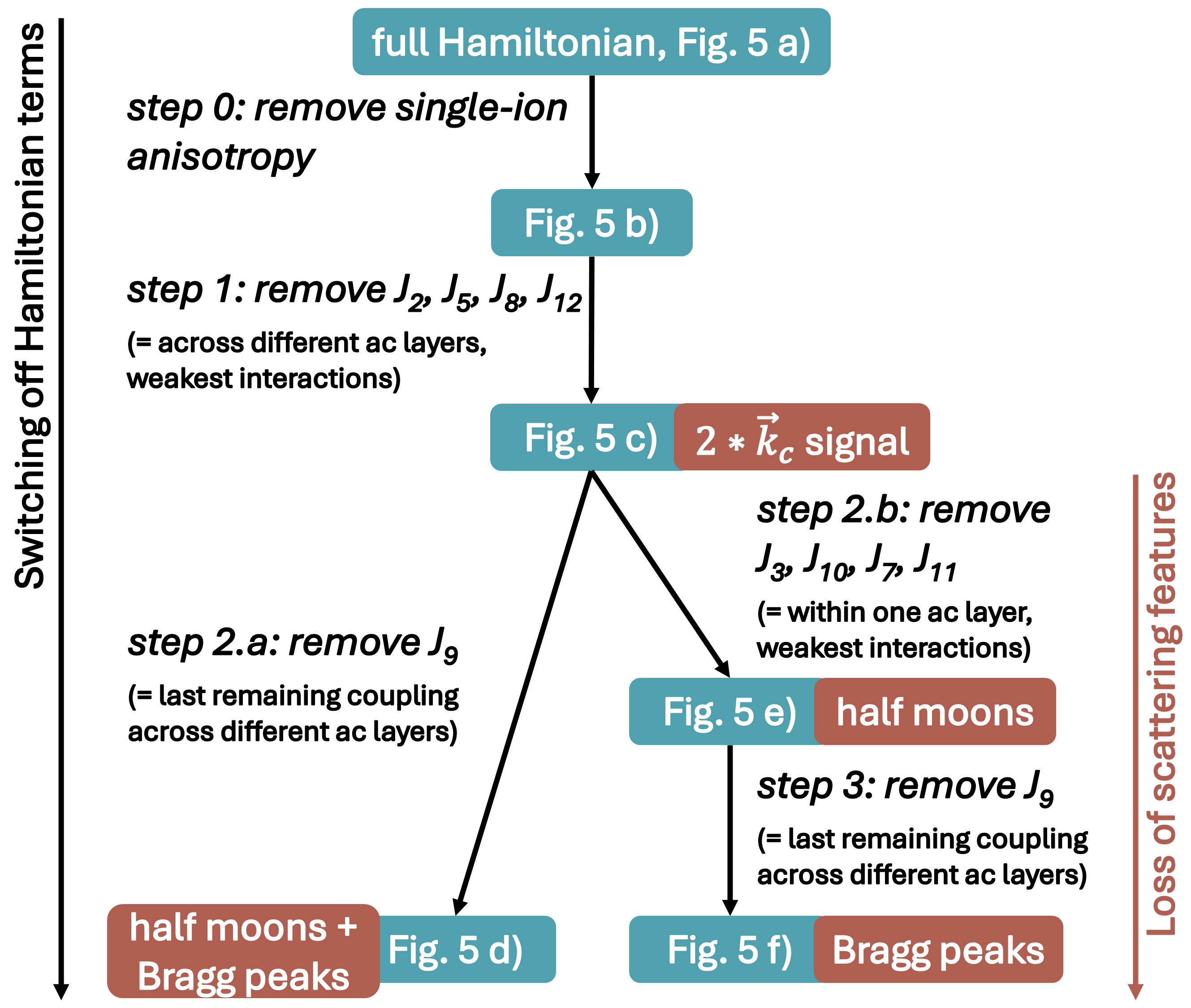}
    \caption[Workflow for simplifying the Hamiltonian]{Our procedure for determining the relevant terms in the Hamiltonian. We start with the full Hamiltonian (top), and then remove terms in  a sequence of steps. The blue boxes provide the reference to the corresponding scattering map in Fig. \ref{fig: Sq_switchingoff}. The brown boxes indicate which scattering features disappear.}
    \label{fig: switchoffJs_graph}
\end{figure}

\begin{figure}

    \centering
    \includegraphics[width=.95\linewidth]{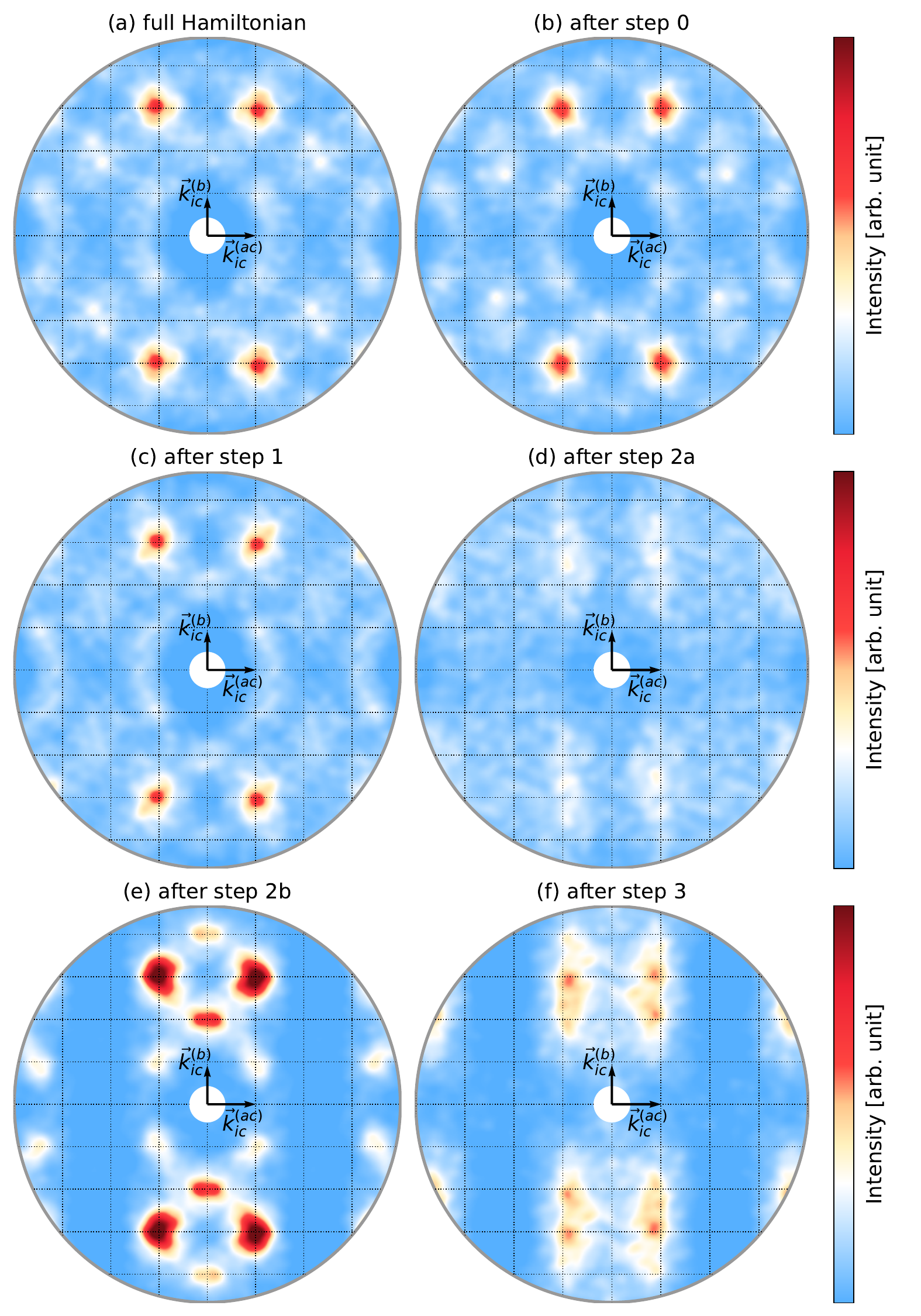}
\caption{Simulated $S(\overrightarrow{q})$, computed with a $36\times24\times36$ supercell at 20 K, for different Hamiltonian models as defined in Fig. \ref{fig: switchoffJs_graph}.
}
\label{fig: Sq_switchingoff}
\end{figure}

\indent Next, we identify which interactions give rise to particular scattering signatures, by switching off individual terms in the Hamiltonian in turn and re-calculating $S(\overrightarrow{q})$ at 20 K, similar to Ref.~\cite{tosic_2024}. Our workflow is presented in Fig.~\ref{fig: switchoffJs_graph}.\\
\indent We start with the full Hamiltonian and remove the single-ion anisotropy (\textit{step 0} in Fig. \ref{fig: switchoffJs_graph}). The scattering maps with the anisotropy term (Fig.~\ref{fig: Sq_switchingoff}\,a)) and without it (Fig.~\ref{fig: Sq_switchingoff}\,b)) show the same features, indicating that the anisotropy is not responsible for the short-range correlations in the paramagnetic regime, which instead must be caused by the exchange interactions. This finding is consistent with the conclusions of Ref.~\cite{lautenschlager_magnetic_1993}, that exchange interactions dominate at higher temperatures and give rise to the incommensurate AF2 and AF3 spiral orders, whereas the anisotropy prevails at low temperatures and gives rise to the collinear AF1 phase.\\
\indent Next, we remove the four weakest exchange couplings, $J_2$, $J_5$, $J_8$ and $J_{12}$, which all couple across different $ac$ layers of Mn zigzag chains (\textit{step 1} in Fig. \ref{fig: switchoffJs_graph}). Fig.~\ref{fig: Sq_switchingoff}\,c) shows that the Bragg peaks and the half moons persist with this simpler Hamiltonian, whereas the $2*\vec{k}^{\,}_{c}$ signals disappear from our calculated structure factor, indicating that the $2*\vec{k}^{\,}_{c}$ signals are a result of the delicate balance between all 12 exchange interactions.\\
\indent We can further simplify the Hamiltonian in two distinct ways: either removing the last remaining coupling across different $ac$ layers - $J_9$ - (\textit{step 2.a} in Fig. \ref{fig: switchoffJs_graph}) or reducing the frustration within the $ac$ plane by switching off $J_3$, $J_{10}$, $J_7$ and $J_{11}$ (\textit{step 2.b} in Fig. \ref{fig: switchoffJs_graph}). We find that the computed $S(\overrightarrow{q})$ for both approaches now lack the  dispersive half moons (Figs. \ref{fig: Sq_switchingoff}\,d) and\,e) ). Our calculations show that by removing the frustrating interactions across different $ac$ layers (\textit{step 2.a}) or by suppressing the interaction within a given $ac$ layer (\textit{step 2.b}), the half moons disappear. Therefore, we conclude that the half-moon shaped correlated diffuse scattering is a result of frustration both within and across $ac$ planes.\\
\indent Interestingly, these two approaches do not influence the Bragg peak signal in the same way. A Hamiltonian without $J_3$, $J_{10}$, $J_7$ and $J_{11}$ continues to reproduce the magnetic Bragg peak signals (Fig. \ref{fig: Sq_switchingoff}\,e) ), and only results in their suppression if $J_9$ is subsequently additionally switched off (\textit{step 3} in Fig. \ref{fig: switchoffJs_graph}) (see Fig. \ref{fig: Sq_switchingoff}\,f) ). This is not the case when only switching off $J_9$ (\textit{step 2.a}), which completely suppresses the magnetic Bragg peaks. We conclude that $J_9$, being the dominant coupling across different $ac$ layers, is vital to the onset of long range magnetic order.\\

\clearpage

\section{Discussion and Outlook}

\indent In summary, we find that a Heisenberg model with competing exchange interactions up to the 12\textsuperscript{th} nearest neighbor captures the paramagnetic half-moon shaped correlated diffuse scattering in MnWO\textsubscript{4}, including its persistence into the paramagnetic regime, as measured in Ref. \cite{holbein_neutron_2016}.\\
\indent We stress that we achieve agreement with experiment by including only competing symmetric Heisenberg interactions in our Hamiltonian; the Dzyaloshinskii–Moriya interaction \cite{xiao_spin-wave_2016}, single ion anisotropy \cite{lautenschlager_magnetic_1993} and magnetoelastic couplings \cite{biesenkamp_structural_2020} are not needed to model the half moon signatures, despite giving rise to INS signatures caused by electromagnons \cite{xiao_spin-wave_2016} and contributing to the formation of the collinear ground state \cite{lautenschlager_magnetic_1993, biesenkamp_structural_2020}. Furthermore, we note that our model is purely classical, showing that our correlated diffuse scattering arises from classical spin-spin correlations. Interestingly, this type of scattering can also arise from quantum fluctuations, for example in the spin-1/2 triangular-lattice Heisenberg antiferromagnets, where it has been attributed to intermediate quasi-particle ordering states \cite{chen_2019}.\\
\indent We find that the half moons vanish upon removing all frustrating interactions across different $ac$ layers of Mn zigzag chains, as well as upon reducing the frustration within one $ac$ layer. Thus, our results indicate that frustration is a key player behind the appearance of the half moons, despite the absence of the usual geometric frustration found in pyrochlore and triangular systems.\\
\indent Moreover, we conclude that the half moons in MnWO\textsubscript{4} are caused by couplings in all three dimensions. Curved correlated diffuse scattering is also measured in MnO and spinel compounds such as ZnCr$_2$O$_4$ and MgCr$_2$O$_4$, which are well described by models that include couplings between spins or weakly interacting clusters of spins across all three lattice dimensions \cite{hohlwein_magnetic_2003, paddison_magnetic_2018, lee_emergent_2002,bai_magnetic_2019,tomiyasu2013emergence}. In contrast, the rod-like diffuse scattering in h-YMnO\textsubscript{3} and NaFe(WO\textsubscript{4})\textsubscript{2} is due to correlations in two-dimensional planes \cite{holbein_neutron_2016, janas_classical_2021}. Therefore, we suggest that the directionality of the competing exchange interactions in two or three lattice dimensions determines whether the correlated diffuse scattering has a rod-like or half-moon shape, respectively.\\
\indent Our findings are consistent with previous studies on other frustrated, classical spin systems, such as pyrochlores \cite{udagawa_out--equilibrium_2016, mizoguchi_clustering_2017, tomiyasu_molecular_2008, lee_emergent_2002} and MnO \cite{paddison_magnetic_2018}, which show paramagnetic correlated diffuse scattering associated with cluster formation. Future studies, such as reverse Monte Carlo calculations \cite{paddison_magnetic_2018} and energy-resolved $S(\overrightarrow{q},w)$ measurements, could unveil more details on the microscopic arrangement of spins associated with the correlated diffuse signatures in MnWO\textsubscript{4} since signals from clusters of spins are frequency independent, whereas spin waves appear at well-defined frequencies \cite{janas_classical_2021}. We also suggest studying the effect of an applied magnetic field on the correlated diffuse scattering, which would provide insight into any net magnetic moment associated with the correlated regions.\\
%OUTLOOK
\indent While our study suggests that three-dimensional, magnetically-frustrated AFM interactions are at the origin of the half moons in MnWO\textsubscript{4}, several open questions remain. Do the half moons persist below $T_\mathrm{N}$, and what is their energy dependence? Are the short-range correlations associated with the correlated diffuse scattering a prelude to the onset of long-range order, or are these two phenomena independent, as suggested for h-YMnO\textsubscript{3} \cite{janas_classical_2021,sato2003unconventional}?
Since quantum systems can show the same type of scattering features as MnWO\textsubscript{4}, how can one differentiate between classical and quantum signatures in diffuse scattering?
We hope that our work inspires future theoretical and experimental investigations on the interplay between correlated diffuse scattering and magnetic phase transitions, in both classical and quantum systems.\\

\textit{Acknowledgements.} We thank Simon Holbein and Yinguo Xiao for making their original data available to us, and Pascale Deen for invaluable feedback on our results. This work was funded by ETH Zürich and the European Research Council (ERC) under the European Union’s Horizon 2020 research and innovation program project HERO grant agreement No. 810451. The Swiss National Supercomputing Center provided us with computational resources under project ID s1128.

\providecommand{\noopsort}[1]{}\providecommand{\singleletter}[1]{#1}%

\cleardoublepage

\begin{figure*}[!h]
\section*{Supplementary Information}
    \centering
    \begin{subfigure}{1\linewidth}
        \centering
        \includegraphics[width=0.8\linewidth]{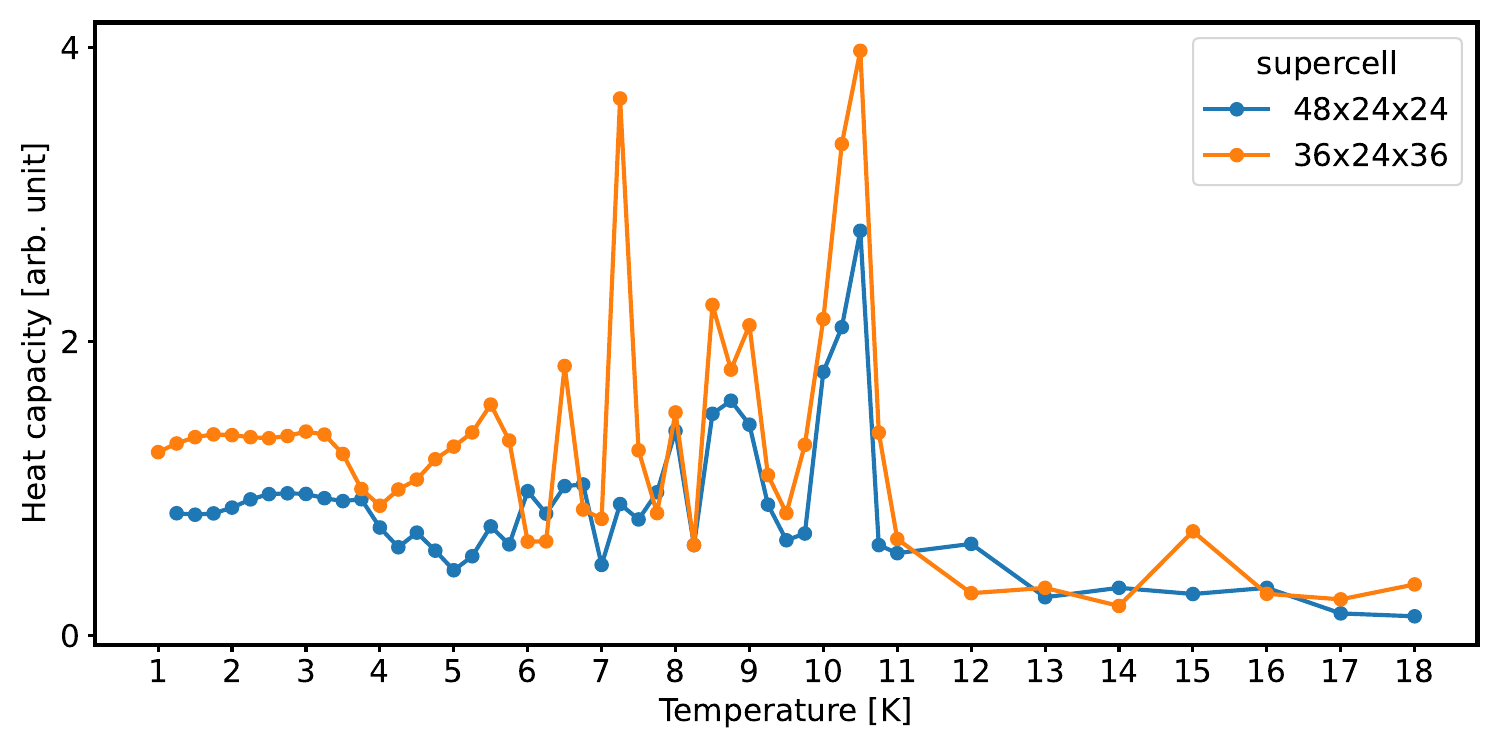}
        \label{fig:heatCapacity_supercellCasestudy}
    \end{subfigure}
    
    \vspace{1em} 
    
    \begin{subfigure}{1\linewidth}
        \centering
        \includegraphics[width=0.8\linewidth]{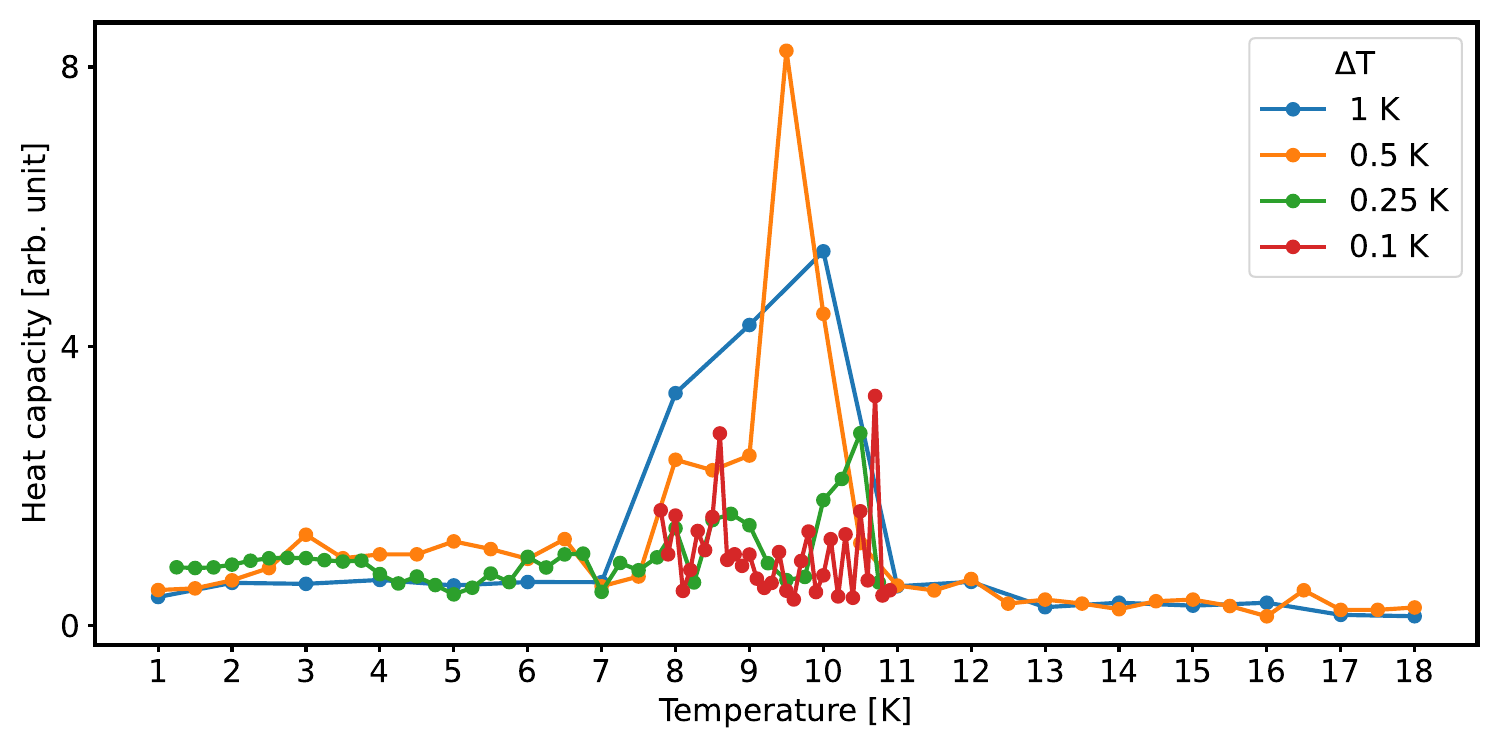}
        \label{fig:heatCapacity_temperatureIntervals}
    \end{subfigure}
    
    \caption{Calculated heat capacity, at the end of the measurement phase, for different values of the cool down speed and supercell size. (a) The 36$\times$24$\times$36 and 48$\times$24$\times$24 supercells at 0.25 K temperature steps, and (b) the 36$\times$24$\times$36 supercell at 1 K, 0.5 K, 0.25 K and 0.1 K temperature steps.}
    \label{fig:combined_heatCapacity}
\end{figure*}

\begin{figure*}[htbp]
    \centering

        \begin{subfigure}[b]{0.5\textwidth} 
        \centering
        \includegraphics[height=5cm]{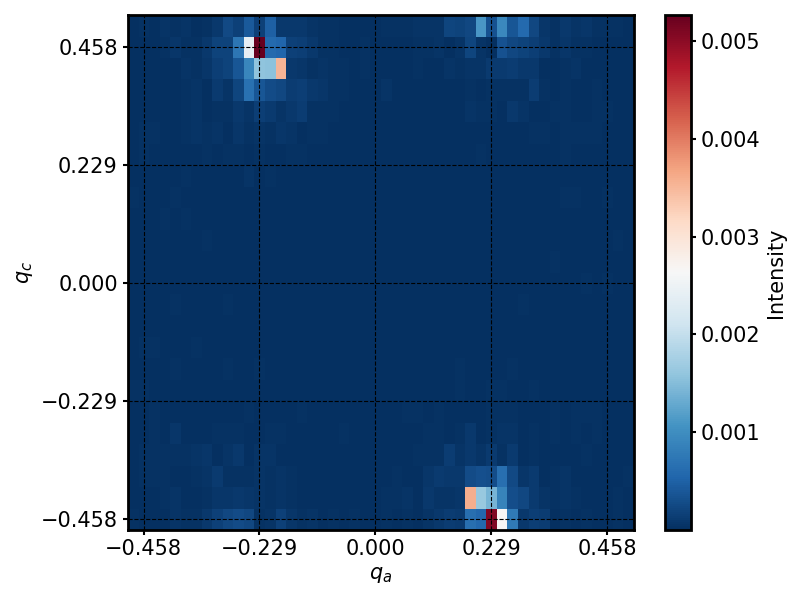}

    \end{subfigure}%
    \hfill
    \begin{subfigure}[b]{0.5\textwidth} 
        \centering
        \includegraphics[height=5cm]{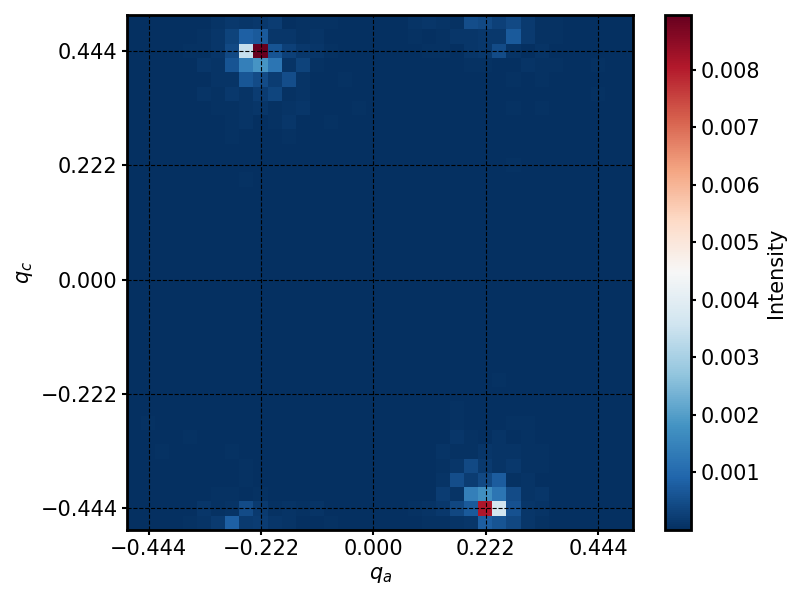}

    \end{subfigure}

    \begin{subfigure}[b]{0.5\textwidth} 
        \centering
        \includegraphics[height=5cm]{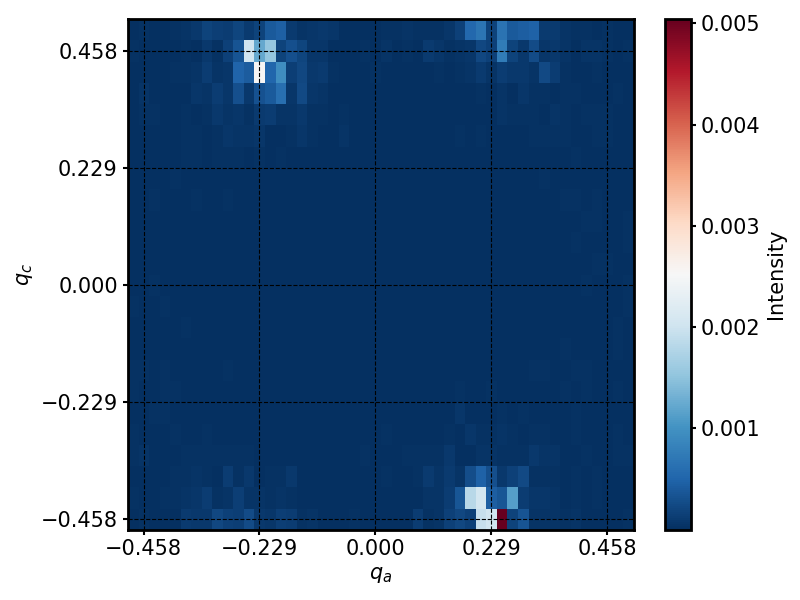}

    \end{subfigure}%
    \hfill
    \begin{subfigure}[b]{0.5\textwidth} 
        \centering
        \includegraphics[height=5cm]{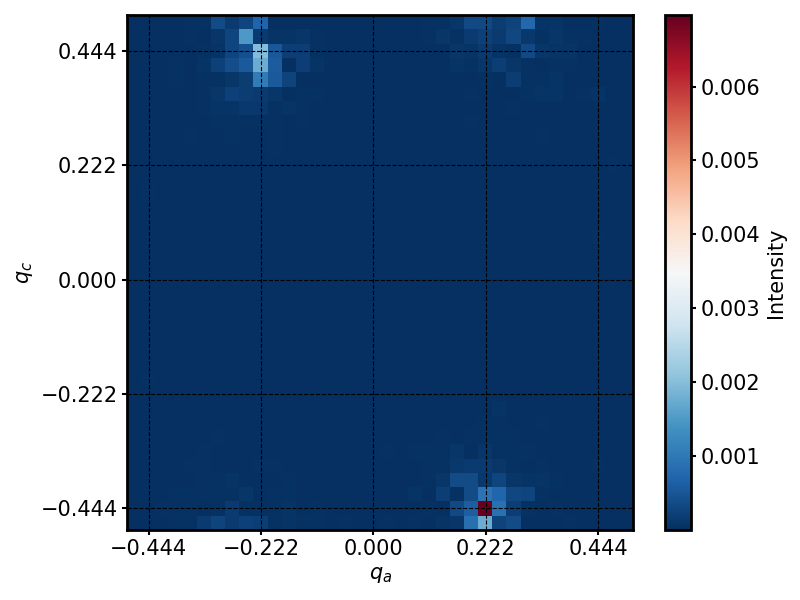}

    \end{subfigure}
    \caption[Static structure factor calculations in the full Brillouin zone]{Computed $S(\overrightarrow{q})$ at 11 K in the full Brillouin zone, with the $q$ values given in fractional coordinates. Left: 48$\times$24$\times$24 supercell, right: 36$\times$24$\times$36 supercell, at $q_b = 0.5$ (top) and $q_b = 0.458$ (bottom). All other slices with $q_b < 0.5$ and $q_b > 0.5$ have lower intensities.
    }
    \label{fig:Sq_fullcellSlices}
\end{figure*}

\begin{figure*}

\subfloat{\includegraphics[width=.7\linewidth]{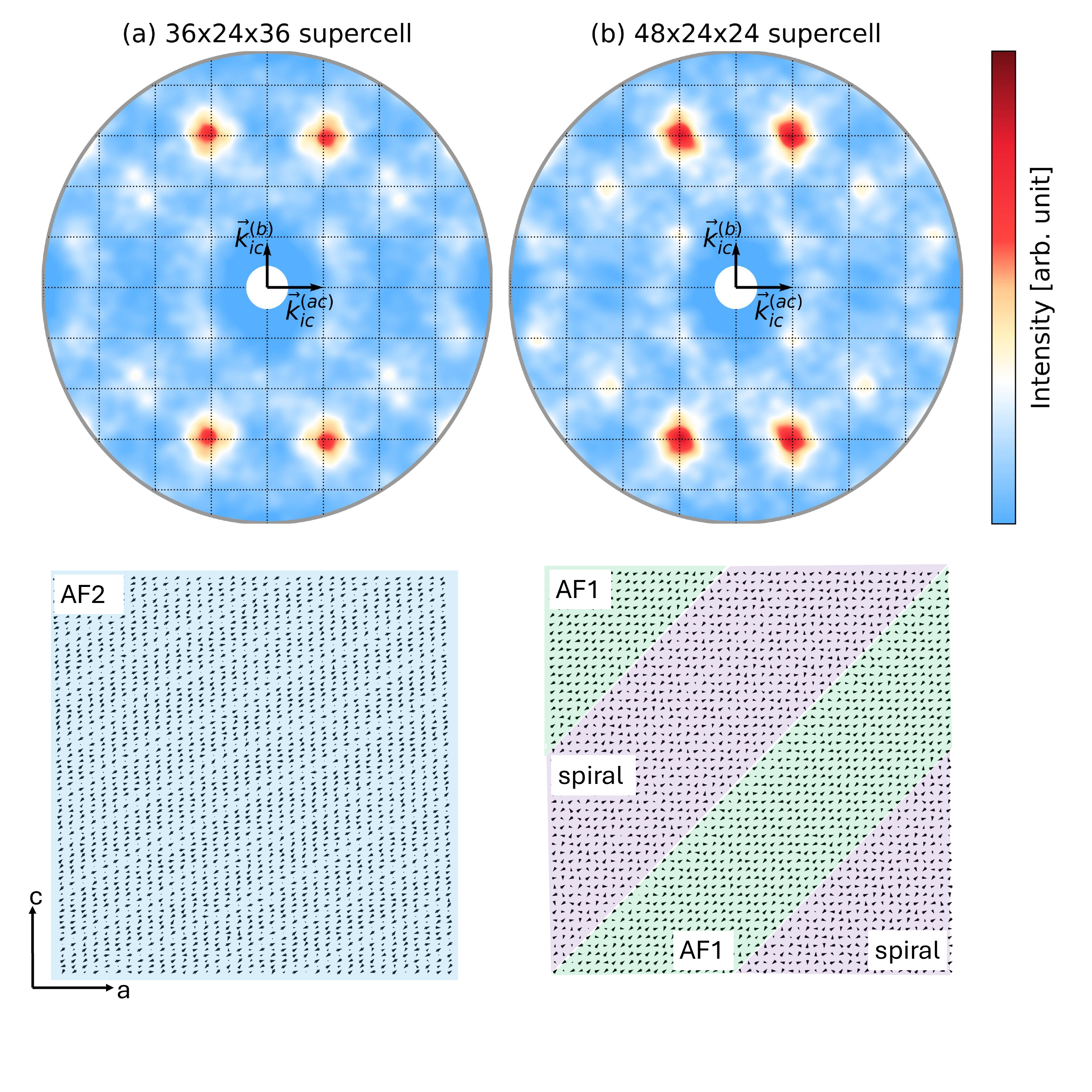}}
\caption{Comparison of the (a) 36$\times$24$\times$36 and (b) 48$\times$24$\times$24 supercells. Top panel: $S(\overrightarrow{q})$ at 20 K, with (a) $\vec{k}^{\,(ac)}_{ic} = [-0.222, 0, 0.444]$ and (b) $\vec{k}^{\,(ac)}_{ic}  = [-0.229,$ $ 0, 0.458]$. Bottom panel: Simulated magnetic configurations at 1 K, plotted in the $ac$ plane projection. The orientations of the magnetic moments alternate for subsequent unit cells in the $b$ direction, consistent with $\vec{k}^{\,}_{c}$ and $\vec{k}^{\,}_{ic}$ in experiment. The configurations are obtained upon cooling from the paramagnetic regime, as described in the Methods section of the main text. }
\label{fig:scscizes}
\end{figure*}

\begin{figure*}
    \centering
    \includegraphics[width=0.5\linewidth]{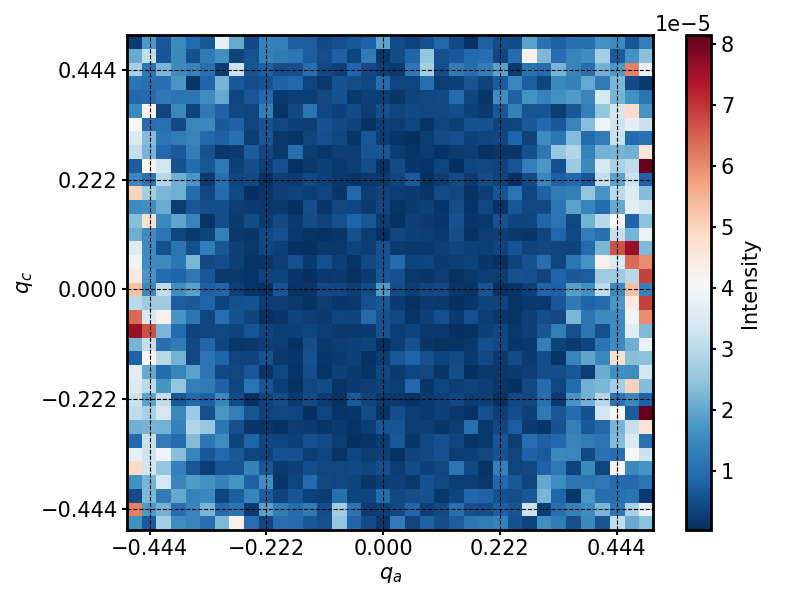}
    \caption{Computed $S(\overrightarrow{q})$ at 11 K and $q_b = 0$ for the 36$\times$24$\times$36 supercell, with the $q$ values given in fractional coordinates.}
    \label{fig:zeroslice}
\end{figure*}

\end{document}